\documentclass[proof]{WileyASNA-v1}

\articletype{Article Type}%

\begin{document}

\title{On the True Nature of the Contact Binary CRTS J192848.7-404555}

\author[1]{Surjit S Wadhwa*}

\author[1]{Ain Y DeHorta}

\author[1]{Miroslav Filipović}

\author[1]{Nick F H  Tothill}

\authormark{WADHWA \textsc{et al}}

\address[1]{\orgdiv{School of Science}, \orgname{Western Sydney University}, \orgaddress{\state{NSW}, \country{Australia}}}



\corres{*Surjit Wadhwa, \email{19899347@student.westernsydney.edu.au}}


\abstract{CRTS J192848.7-404555 was recognised as a potential contact binary merger candidate on the basis of survey photometry analysis. We have carried out follow up ground based photometry of the system and show that at the recorded coordinates for the system there are two stars approximately 3 seconds of arc apart. Our analysis shows that the fainter of the two stars is the actual variable while the slightly brighter star is of fixed brightness. In addition we show that the reported survey photometry is the result of both stars being treated as a single light source with resultant erroneous light curve solution. The true nature of CRTS J192848.7-404555 shows it to be a low mass contact binary system with a high mass ratio of 0.425, high amplitude of 0.69 magnitude and shallow 24\% contact. The system does not have features of orbital instability and is not a potential merger progenitor.}




\maketitle


\section{Introduction}\label{sec1}

CRTS J192848.7-404555 (C1928) ($\alpha_{2000.0} = 19\ 28\ 48.884$, $\delta_{2000.0} = -40\ 45\ 54.539$) (= ASASSN-V J192848.87-404554.0, 1SWASP J192848.72-404554.6, ASAS J192849-4045.9) was recognised as a contact binary system by the Catalina Real-time Transient Survey (CRTS) \citep{2017MNRAS.469.3688D} with a period of 0.320028d, brightest V band magnitude of 12.04 and V band amplitude of 0.27 magnitudes. Catalina magnitudes were converted to V band as per The Catalina Surveys Data Release 2 notes available online at http://nesssi.cacr.caltech.edu/DataRelease/ (accessed 10/10/2022). Other surveys such as the All Sky Automated Survey - Super Nova (ASAS-SN) \citep{2014ApJ...788...48S, 2020MNRAS.491...13J}, SuperWASP \citep{2021MNRAS.502.1299T} and All sky Automated Survey \citep{2002AcA....52..397P} have all reported similar basic parameters. The light curves from these surveys are illustrated in Figure 1. 

\begin{figure*}[!ht]
    \label{fig:A1928F1}
	\includegraphics[width=\textwidth]{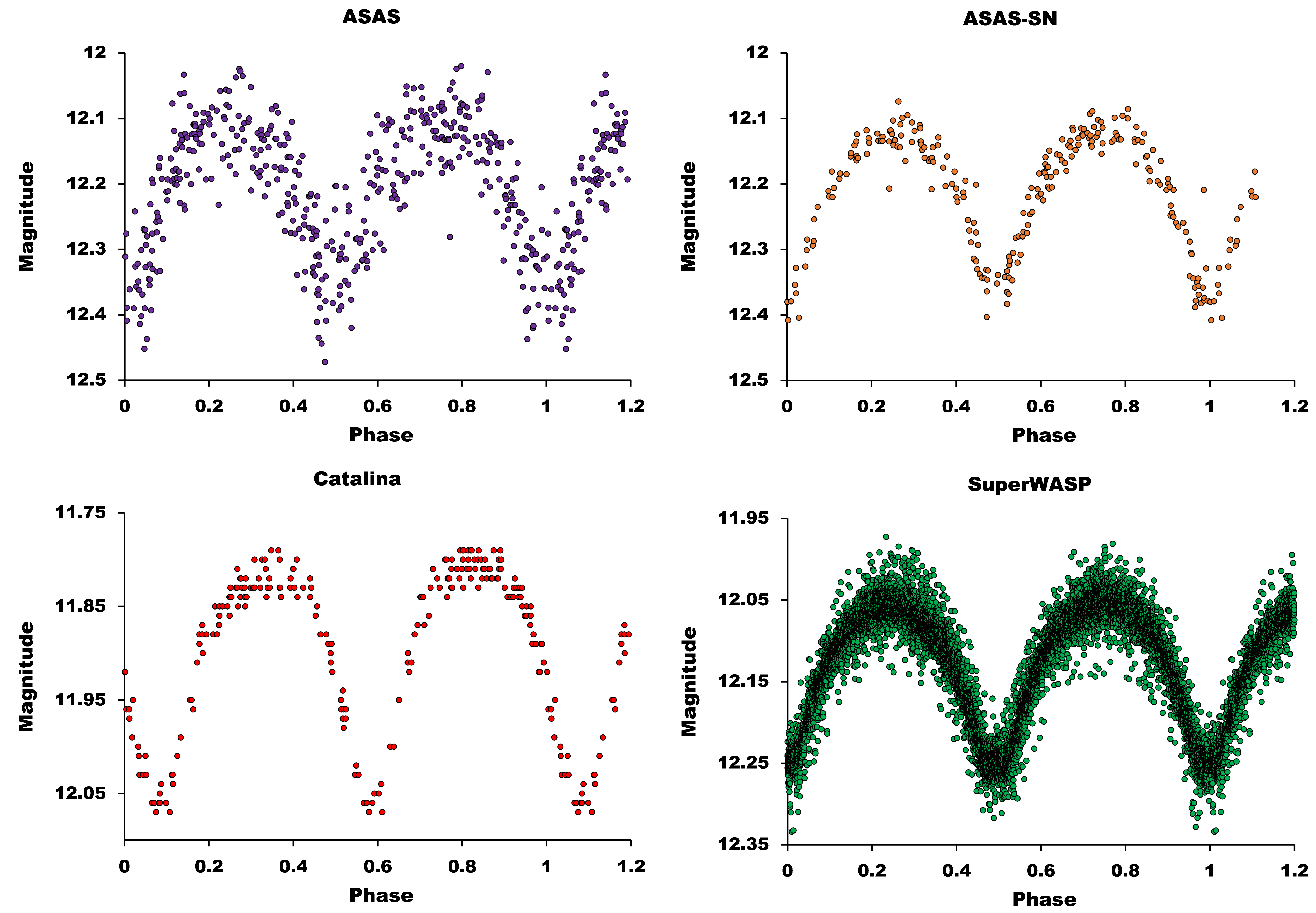}
    \caption{Light curves of C1928 from 4 sky surveys as labelled. All show a bright contact binary system light curve with amplitude near 0.3. There is slight variation in the actual magnitude values is due to the different band pass for each survey. The ASAS and ASAS-SN surveys use standard V-Band filters, Catalina survey data is unfiltered while the SuperWASP survey employs a broadband filter from 400 to 700 nm}
    \end{figure*}

The SuperWasp photometry of C1928 was modelled by \citet{2022arXiv220800626W} and indicated a very low mass ratio ($q=0.08$), low inclination ($i=67^{\circ}$) system in marginal contact. As part of our ongoing project to identify and confirm low mass ratio contact binary systems, in this study we present the results and analysis of ground based photometry of C1928 and show that the survey photometry is blended from a nearby star approximately 3 seconds of arc roughly west of the contact binary system. The contact binary system is significantly fainter than reported with a much larger amplitude and significantly higher mass ratio. As would be expected physical parameters such as masses, radii and separation of the components are also different to those derived from analysis of the survey photometry.

\section{Observations}\label{sec2}
C1928 was observed with the 0.4m telescope network of the Las Cumbres Observatory (LCO) over 4 days in June 2022. Full cycle Bessel V band photometry was acquired with 366 observations in total with typical exposure time of 45 seconds. In addition concurrent 50 Bessel B band observations (exposure time 60 seconds) during times of minima were acquired to document the (B-V) colour of the system. Review of the initial V band images (Figure 2) clearly shows two stars approximately 3 seconds of arc apart at the recorded location of C1928. As most survey data is obtained either with a telephoto lens or small aperture telescopes, we postulated there maybe some that there was a potential for mis-identification and or blending of the survey light curves and the true nature of C1928 may differ from that described based on published survey photometry. The main aim of this study was to identify which of the two stars at the reported location was the variable and once identified to fully elucidate the nature of the likely contact binary system.

\begin{figure}
    \label{fig:A1928F2}
	\includegraphics[width=\columnwidth]{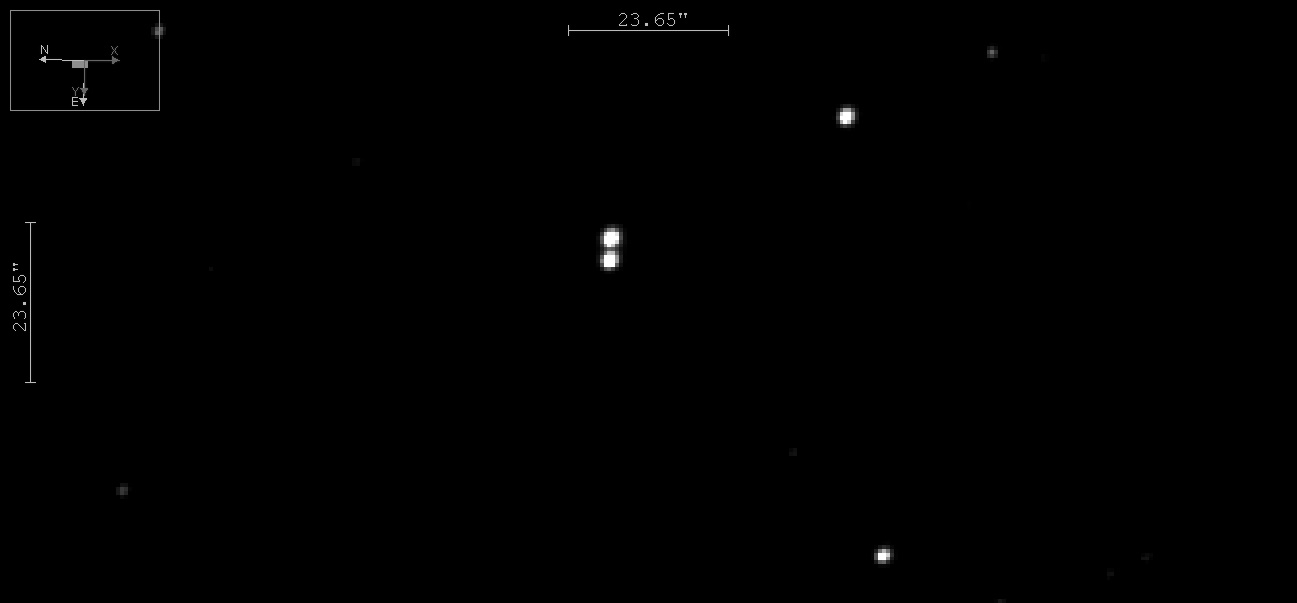}
    \caption{At the location of C1928 there are clearly two stars one slightly brighter.}
    \end{figure}

We performed aperture photometry on both stars using 2MASS-19284698-4046286 (V = 13.29, B = 14.27) as the comparison star and 2MASS-19285265-4046355 as the check star. We used the AAVSO Photometric All-Sky Survey \citep{2015AAS...22533616H} calibrations for the comparison star magnitudes. The data was folded using the revised period (see below) and normalised to the brightest magnitude. The V band light curve of the slightly brighter (upper star in Figure 2) is essentially constant with V magnitude of 12.76. There is possibly some minor variability when compared to the check star however it is likely to be less than 0.1 magnitude in amplitude. We could not find any short term periodicity. The V band light curve of the fainter (lower star in Figure 2) shows a classical contact binary light curve however the amplitude, unlike survey photometry, is considerably larger at 0.69 (V: 12.86 - 13.55) magnitude. In other respects the light curve has similar shape to the survey light curves. To confirm our suspicion that the survey light curves were the result of full contamination by the non variable star we performed photometry with both stars acting as a single light source. The resulting light curve is very similar to the survey light curves with a maximum brightness V magnitude of 12.08 and amplitude of 0.29 magnitudes. All three light curves and the check star curve are illustrated in Figure 3.

\begin{figure}
    \label{fig:A1928F3}
	\includegraphics[width=\columnwidth]{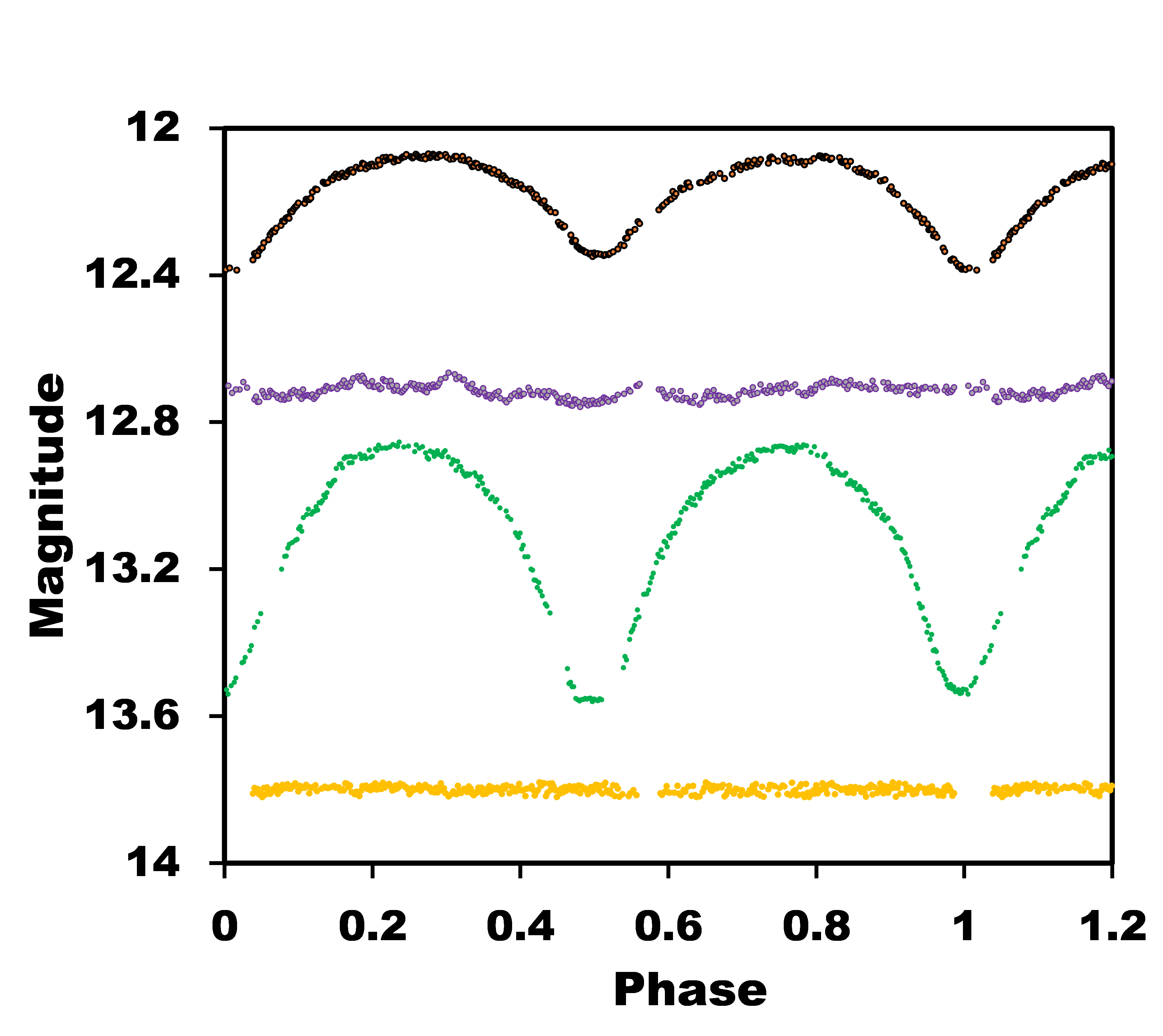}
    \caption{Light curves of the two stars at the location of C1928. Top curve (black) illustrates the light variation when both stars are taken as a single light source. The middle (purple) curve is that of the bright star while the bottom (green) curve is of the fainter star. The yellow curve represents the check star}
    \end{figure}

Based on the times of minima observed we revised the orbital elements as follows:\\
\begin{center}
    
    $HJD_{min} = 2459753.121103(\pm298) + 0.3200310(\pm233)E$\\                   

\end{center}

The B-V value was the same at both the primary and secondary eclipses at 0.77 which is quite different to 0.65 reported with the ASAS-SN survey. We note that the B-V value of the pair taken as a single light source is 0.67, similar to the value reported by ASAS-SN again confirming that the survey photometry data includes both stars.

\section{Light Curve Analysis}
Given the presence of complete eclipses, light curve analysis without radial velocity data is feasible and should yield accurate results for the geometric parameters \citep{2005Ap&SS.296..221T}. We used the 2009 version of the Wilson-Devenney code to analyse the V band photometry data. As there is no appreciable O'Connell effect only unspotted solutions were modelled. The accepted grid/q search method was used to obtain the mass ratio ($q$). The effective temperature of the primary ($T_1$) is usually fixed during the search procedure. As the recent $Gaia$ EDR 3 separates the two stars \citep{2022A&A...658A..91A, 2022arXiv220800211G} we estimated $T_1$ as 5900K. As usual gravity darkening coefficients were fixed as ($g_1 = g_2 = 0.32$), bolometric albedo was fixed at ($A_1 = A_2 = 0.5$) and simple reflection treatment was applied. Limb darkening coefficients were interpolated from \citet{1993AJ....106.2096V}. The mass ratio search grid along the observed and fitted light curves are illustrated in Figure 4 and the light curve solution summarised in Table 1.

\begin{figure*}
    \label{fig:A1928F4}
	\includegraphics[width=\textwidth]{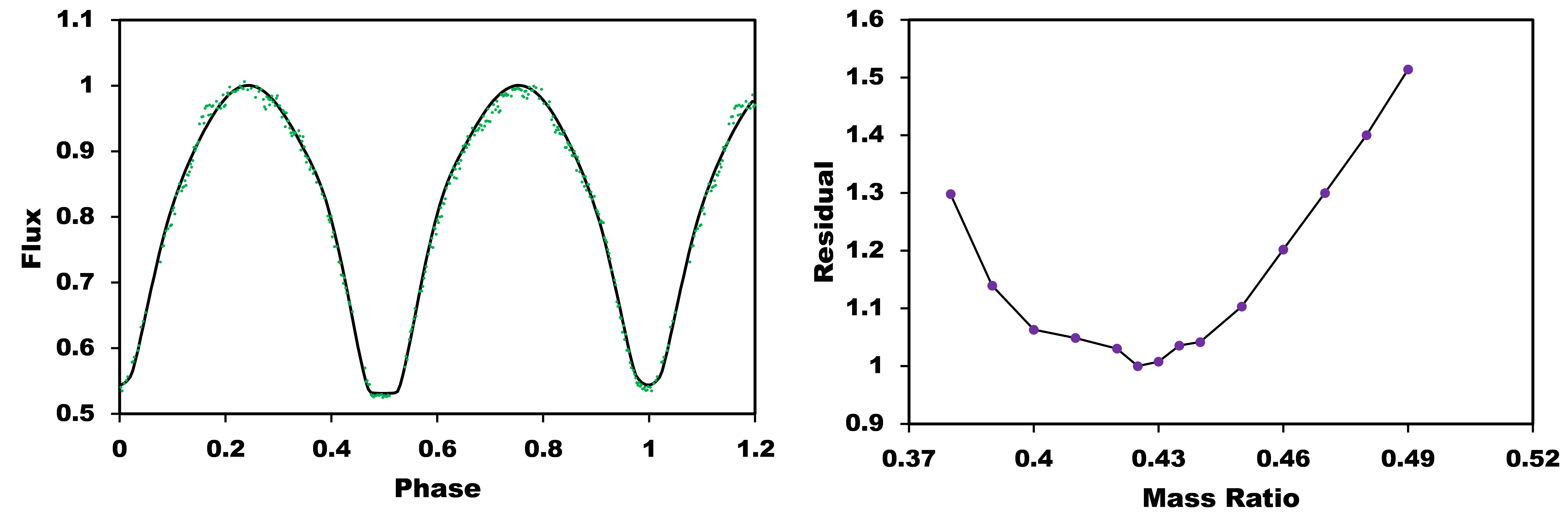}
    \caption{Observed (green) and fitted (black) light curves are illustrated on the left while the mass ratio search grid on the right. The search grid residuals were normalised to the minimum value and only the portion near the minimum residual is illustrated for clarity}
    \end{figure*}

The system is observed edge on ($i=90^0$) and found to have a slightly hotter secondary ($T_2 = 6148K$) with a mass ratio of 0.425 and shallow fillout of 24\%. The light curve solution is very different to the solution based on survey data which suggested a very low mass ratio of 0.08 with low inclination and marginal contact.

\section{Physical Properties}
Apart from the mass ratio the next most critical parameter required to estimate other physical characteristics is the mass of the primary. It is well established that the primary component of a contact binary systems follow in general a main sequence profile \citep{2013MNRAS.430.2029Y}. We estimate the mass of the primary ($M_1$) as the mean of combined distance and color based calibrations.

As there is a risk of contamination with published color magnitudes of the system we first determined the distance and reddening corrected $(B-V)_0$ value of the system as follows: Firstly, we determined the line of sight reddening at infinity $E(B-V)_{\infty}$ centered on the system coordinates using the \citet{2011ApJ...737..103S} dust maps. The value was then scaled to the $Gaia$ distance ($461.4 \pm10 pc$) \citep{2022A&A...658A..91A} $E(B-V)_d$ using the equation \citep{2008MNRAS.384.1178B}:

\begin{equation}
    E(B-V)_d = E(B-V)_{\infty}\Bigg[1-exp\bigg(-\frac{|dsinb|}{h}\bigg)\bigg]
\end{equation}
In the equation $d$ is the $Gaia$ distance, $b$ is the galactic latitude of the system and $h$ is the galactic scale height, taken as $h=125pc$ as per \citet{2008MNRAS.384.1178B}. The distance and reddening corrected $(B-V)_0$ was determined as 0.70. 

As the $Gaia$ mission does separate the two stars we also make use of the $Gaia$ color $(Bp-Rp) = 0.94$ corrected to $(Bp-Rp)_0=0.82$ using the $Gaia$ line of extinction of $E(Bp-Rp)_{\infty} = 0.13$ and $E(Bp-Rp)_d = 0.12$. 

In addition to the two color estimates we estimated the luminosity of the primary based on the $Gaia$ distance and our photometry data. The eclipses are of near equal depth thus the apparent magnitude at mid eclipse (13.53) represents the apparent magnitude of the primary component. Given the distance and the distance corrected line of sight extinction determined above we calculate the absolute magnitude of the primary ($M_{V1}$) as $5.42\pm 0.05$. From the difference in peak and eclipse brightness we estimate the absolute magnitude of the secondary ($M_{V2}$) as $5.59\pm0.05$. 

We used the April 2022 updated calibration tables of \citet{2013ApJS..208....9P} for low mass ($0.6M_\odot < M < 1.5M_\odot$) stars along with $(B-V)_0$, $(Bp-Rp)_0$ and $M_{V1}$ to interpolate three different values for the mass of the primary as $0.95M_\odot,$ $ 1.01M_\odot$ and $0.91M_\odot$ respectively. We adopt the mean of these as the mass of the primary ($M_1 = 0.96\pm 0.01M_\odot$). We use the error based on the distance estimate as this was of the highest order. From the mass of the primary we estimate the mass of the secondary ($M_2$) as $0.41\pm 0.02 M_\odot$. Other physical parameters such as the radii of the components ($R_{1,2}$), the separation ($A$) and density ($\rho_{1,2}$) can be estimated from the fractional radii of the components (from light curve solution), mass ratio and Keplers' third law as previously described \citep{2022JApA...43...42W}. Errors were propagated from the errors in the mass of the components by far the greatest contributors to overall error. Physical parameters are summarised in Table 1.

\citet{2021MNRAS.501..229W} introduced simplified quadratic relations linking the mass of the primary of contact binary systems with the mass ratio at which orbital instability is likely. Based on those relations one can estimate a narrow range of the mass ratio at or below which orbital instability is likely. \citet{2022arXiv220800626W} when analysing the survey photometry of C1928 estimated the mass of the primary based on the main sequence calibration of the J-H magnitude as $1.14M_\odot$ yielding an instability mass ratio range of 0.082 - 0.093. As noted above their analysis of the survey photometry suggested a mass ratio of 0.08 for C1928 and they concluded that the system was likely a merger candidate. The current analysis suggests that the primary is likely significantly smaller with a mass of $0.96M_\odot$ with a resulting instability mass ratio range of 0.107 - 0.126. Our light curve solution indicates a mass ratio considerably higher thus suggesting that the system is likely quite stable and unlikely to be a merger candidate. Similarly, the estimated separation of the system at $2.18R_\odot$ is significantly higher than the predicted instability separation range of $1.72R_\odot - 1.81 R_\odot$ based on the formulae described in \citep{2021MNRAS.501..229W}.

\section{Energy Transfer and Angular Momentum Loss}
\subsection{Energy Transfer and Density}
The atypical nature of the secondary component of contact binary systems has long been recognized \citep{1948AnAp...11..117S}. In particular, the secondary is brighter and larger than the main sequence counterparts. \citet{1968ApJ...151.1123L} suggested that the discrepancy in the brightness and size of the secondaries was due to the transfer of energy from the primary to the secondary within the common envelope. A number of authors have explored the relationship between the mass ratio and luminosity and the transfer of energy between the components \citep{1994ApJ...434..277W, 2004A&A...426.1001C} with general agreement that energy transfer is a function of both the mass ratio and luminosity ratio. 

\citet{2004A&A...426.1001C} introduced the energy transfer parameter defined as:
\begin{equation}
    \beta = \frac{L_{1,Observed}}{L_{1,ZAMS}}
\end{equation}

They showed that the transfer parameter can be estimated as

\begin{equation}
    \beta = \frac{1+q^{4.6}}{1+q^{0.92}\big(\frac{T_2}{T_1}\big)^4}
\end{equation}
and the transfer of the luminosity from the primary to the secondary can be estimated as

\begin{equation}
    \Delta L = (1-\beta)L_1
\end{equation}
 In the equations $q$ is the mass ratio, $T_{1,2}$ the temperatures of the primary and secondary components, $L_1$ is luminosity of the primary in solar units and $\Delta L$ the transferred luminosity in solar units. $ZAMS$ indicates Zero Age Main Sequence.

 We estimate the luminosity transfer from the primary to the secondary to be in the order of $0.22L_\odot\pm0.05$ which is quite significant given the intrinsic luminosity of the primary is in the order of $0.66L_\odot$.

 The effect of energy transfer on the radius of the secondary was explored by \citet{2009MNRAS.396.2176J}. They showed that higher the energy transfer rate the greater the impact on the radius of the secondary. As expected given the high luminosity transfer in the system the estimated radius of the secondary $R_2=0.73R_\odot$ is nearly double the calibrated main sequence star radius. 

 \citet{2001AJ....122..425Y} suggested that the over-luminosity of the secondary is related to the relative higher density of the secondary. The density (in $gcm^{-3}$) of the components can be expressed as a function of the period, radii and mass ratio \citep{1981ApJ...245..650M} and it is easy to show that the difference in densities of the components can be expressed as:
\begin{equation}
    \Delta\rho = \frac {0.0189q}{r_2^3(1+q)P^2} - \frac {0.0189}{r_1^3(1+q)P^2}
\end{equation}
where $q$ is the current mass ratio, $r_{1,2}$ relative radii of the primary and secondary and $P$ is the period in days. $\Delta\rho$ for C1928 was estimated as $0.017\pm0.002gcm^{-3}$ indicating near equal density despite vastly different masses and radii.

\subsection{Angular Momentum Loss}

Most contact binary systems have periods of less than 1 day with short period cutoff at approximately 0.22 days. Variation in period has been suggested as a marker of orbital instability, however, variation is commonly observed with decreasing and increasing variations observed equally \citep{2018MNRAS.474.5199L}. Numerous mechanisms such as light time travel effects, apsidal motion, magnetic activity cycles, mass transfer or loss can result in either shortening or lengthening of the period. Long term period decrease however is usually due to angular momentum loss \citep{2018MNRAS.474.5199L}. C1928 does not have historical high cadence observations to mount a meaningful period study however it is possible to estimate the current potential period decrease due to angular momentum loss (AML). 

\citet{1994ASPC...56..228B} deduced a theoretical constraint on the rate of AML as per the following equation:

\begin{equation}
\begin{split}
 \frac{dP}{dt} \approx 1.1\times 10^{-8}q^{-1}(1+q)^2\times (M_1 + M_2)^{-5/3}k^2\times\\
 (M_1R_1^4 + M_2R_2^4)P^{-7/3} \times 86400
 \end{split}
\end{equation}
where $M_{1,2}$ are masses of the primary and secondary in solar units, $q$ is the mass ratio, $P$ is the period in days, $R_{1,2}$ are the radii of the primary and secondary components, and $k$ is the gyration radius of the primary. The resultant rate change in the period is in seconds  yr$^{-1}$.

We interpolated the value of the gyration radius ($k$) for low mass rotating and tidally distorted stars as described in \citep{2021MNRAS.501..229W}. We estimate AML for the system as $-0.01sec.yr^{-1}$. If AML was the only source of period change then it would be unlikely any change would have been detected during the 30 or so years of survey observations.

\begin{table}

    \centering
           \begin{tabular}{|c|c|c|c|}
    \hline
          & && \\
         Parameter & Value&Parameter&Value\\
         &&&  \\\hline
         $T_1$ (K) (Fixed) & 5900&$M_2/M_{\odot}$ &$0.41\pm0.02$\\
         $T_2$ (K) & $6148\pm12$&$A$/$R_{\odot}$ & $2.18\pm0.01$\\
         Inclination ($^\circ$) & $90\pm1.0$&$R_1/R_{\odot}$ &  $0.98\pm0.02$\\
         Fillout(\%) & $24\pm2$&$R_2/R_{\odot}$ &  $0.74\pm0.02$\\
         $q$ ($M_2/M_1$) & $0.425\pm0.003$&$\rho_1 (gcm^{-2})$ & $1.41\pm0.05$\\
         $M_1/M_{\odot}$ &$0.96\pm0.01$&$\rho_2 (gcm^{-2})$& $1.39\pm0.05$\\
                           \hline
         
                     \end{tabular}
    \caption{Light curve solution and absolute parameters for C1928}
\end{table}

 \section{Summary and Conclusions}

Analysis of survey photometry in the detection of interesting, particularly low mass ratio contact binary systems has received considerable attention recently \citep{2020MNRAS.493.1565D, 2022JApA...43...42W,  2022MNRAS.512.1244C}. Although it is clear that in most cases survey photometric data is suitable and yields acceptable light curve solutions, careful follow-up evaluation of any interesting system is essential as demonstrated by the present study. The current study shows that at least 4 sets of survey data namely, ASAS, ASAS-SN, SuperWASP and CRTS sampled a double star as a single system yielding erroneous estimates of brightness and amplitude of the light curve. When combined with the reported colour estimates, also shown to be erroneous, have previously yielded physical parameter estimations for the system that are far removed from the values presented in this study. Although survey photometric data offers a vast pool data available for analysis and there are numerous examples in the literature of favourable comparisons of survey and dedicated observations yielding comparable results \citep{2022JApA...43...42W, 2021MNRAS.506.4251Z} care must be taken avoid potential sources of error such as observation scatter and blending. Blending is potentially a common problem given the low resolution of many surveys such as ASAS which has an approximate resolution of 23" \citep{2002AcA....52..397P}, ASAS-SN 17"\citep{2014ApJ...788...48S}, CRTS 9" \citep{2017MNRAS.469.3688D} and SuperWASP 60"\citep{2021MNRAS.502.1299T}. Other examples of survey data yielding results not confirmed by other methods include CN Hyi \citep{2009NewA...14..461O} and V1187 Her \citep{2019PASP..131e4203C}. In both cases the survey analysis over-estimated the mass ratio compared to spectroscopic mass ratio in the case of CN Hyi and dedicated ground based observations in the case of V1187 Her.

In the case of C1928 survey data analysis suggested that the system was of extreme low mass ratio and likely merger candidate. The system actually is a high mass ratio system with a large amplitude. The geometric parameters such as the mass ratio, separation and fill-out are far removed from values indicating orbital instability and we conclude that the system is quite stable and not a potential merger candidate. In other respects the system is similar to other contact binary systems having a secondary that is considerably brighter and larger than the corresponding main sequence counterpart of the same mass. Similarly, the density of the secondary is relatively high and near equal to that of the primary.As the temperature of the secondary is higher than that of the primary, technically the system would be classified as a W-Type. As noted by \citet{2004A&A...426.1001C} W-type systems are more commonly seen where the mass ratio is greater than 0.35 and as such C1928 is similar to other stable systems.

The present study serves as a cautionary tale as to the reliance on survey photometry for formal light curve analysis and absolute parameter determination. As part of our long term project to formally observe survey identified low mass ratio contact binary systems we suspect we will find other such examples.


\section*{Acknowledgments}

This work has made use of data from the European Space Agency (ESA) mission
{\it Gaia} (\url{https://www.cosmos.esa.int/gaia}), processed by the {\it Gaia}
Data Processing and Analysis Consortium (DPAC,
\url{https://www.cosmos.esa.int/web/gaia/dpac/consortium}). Funding for the DPAC
has been provided by national institutions, in particular the institutions
participating in the {\it Gaia} Multilateral Agreement.

This research has made use of the SIMBAD database, operated at CDS, Strasbourg, France.








\bibliography{Wiley-ASNA}%



\end{document}